# SIMPLER QUANTUM COUNTING

CHU-RYANG WIE

*Department of Electrical Engineering, University at Buffalo*
*State University of New York, Buffalo, NY 14260*

A simpler quantum counting algorithm based on amplitude amplification is presented. This algorithm is bounded by $O(\sqrt{N/M})$ calls to the *controlled-Grover* operator where $M$ is the number of *marked* states and $N$ is the total number of states in the search space. This algorithm terminates within $\lceil \log_2 \sqrt{N/M} \rceil$ consecutive measurement steps when the probability $p_1$ of measuring the state $|1\rangle$ is at least 0.5, and the result from the final step is used in estimating $M$ by a classical post processing. The purpose of *controlled-Grover* iteration is to increase the probability $p_1$. This algorithm requires less quantum resources in terms of the width and depth of the quantum circuit, produces a more accurate estimate of $M$, and runs significantly faster than the phase estimation-based quantum counting algorithm when the ratio $M/N$ is small. We compare the two quantum counting algorithms by simulating various cases with a different $M/N$ ratio, such as $M/N \geq 0.125$ or $M/N < 0.001$.

## 1 Introduction

Grover's quantum algorithm searches an unstructured database to find $M$ records that satisfy a given criterion in an $N$-element database [1], but needs to meet several requirements to be useful in practice [2]. The quantum search algorithm uses an estimate on the number of *marked* items $M$ to decide an optimum number of iterations of the *Grover operator* for a high probability of measurement success [3]. This number $M$ is usually estimated using the quantum counting algorithm based on quantum phase estimation [3,4]. We present here a simpler method for estimating the number $M$ of the *marked* items by amplitude amplification rather than phase estimation. We base this algorithm on amplitude amplification using a basic quantum circuit with a single-qubit measurement register, with the final goal at achieving a measurement probability of at least *0.5* for the *|1⟩* state by consecutively increasing the number of *controlled-Grover* iterations, followed by a classical post-processing once the quantum algorithm terminates. This algorithm is simple and reliable, produces a more accurate estimate and has a less demand for the quantum resources than the algorithm based on the phase estimation. Our approach is similar to the work of Svore et.al. [5] in which they improved the performance of the quantum phase estimation algorithm, in that they also used a basic quantum measurement circuit and a classical post-processing.

The quantum counting algorithm [4], based on the *phase estimation algorithm* (*PEA*) and the quantum Fourier transform (*QFT*) algorithm, requires $O(\sqrt{N/M})$ *controlled-Grover* iterations, and thus $O(\sqrt{N/M})$ oracle calls, in order to estimate the phase angle $\theta$, for example, to the $m=\lceil \log \sqrt{N} \rceil + 1$ bits of accuracy [3]. This algorithm requires additional qubits in the measurement register, in order to achieve a high probability of measurement success [3]. The resolution in the phase angle $\theta$ determines the resolution in $M$ via $M=N\sin^2(\theta/2)$. In this *PEA*-based counting algorithm, $M$ is determined by measuring the bits that represent $\phi$ where $\theta=2\pi\phi$. The measurement accuracy depends on the number of qubits in the measurement register as it determines the resolution of $\phi$. When this number $M$ is small relative to the search space of $N$ items, the *PEA*-based counting algorithm can encounter a practical difficulty in both the circuit width and depth because the total number of qubits available is usually limited and the circuit depth grows exponentially with every added qubit to the measurement register in order to achieve an acceptable resolution in the phase angle $\phi$. A certain minimum number of bits for $\phi$ is required in order to achieve an acceptable accuracy for $M$ and $\theta$. On the other hand, our simpler counting algorithm requires a single-qubit measurement register on top of the Grover search circuit and a classical post-processing produces an estimate for the number of marked items.

Our simpler counting algorithm based on the amplitude amplification is performed by consecutively raising $k$, a parameter that determines the number $2^k$ of *controlled-Grover* iterations, in order to boost the measurement probability of the state $|1\rangle$ on the measurement register to a final target value of $0.5$ or higher. This algorithm terminates at or before the $\lceil \log_2 \sqrt{N/M} \rceil^{th}$ measurement step and is followed by a classical post-processing to estimate $M$. In this algorithm any guessing about the needed number of the phase bits is not necessary unlike the *PEA*-based algorithm. We can estimate $M$ and $\theta$ reliably with a lower demand for the quantum resources compared with the *PEA*-based algorithm. In the applications to problems where $M/N$ is very small, this algorithm can run significantly faster than the *PEA*-based algorithm because, as will be shown later, this algorithm halts at or before the $\lceil \log_2 \sqrt{N/M} \rceil^{th}$ measurement step. On the other hand, the *PEA*-based counting algorithm requires $t=\lceil m+\log_2(2+1/2\varepsilon) \rceil$ qubits in the measurement register for an $m$-bit approximation to the phase bits [3]. Where, $1-\varepsilon$ is the probability for a successful measurement in the phase estimation algorithm. Although the difference between the values of $k$ and $t$ may only be 2 or 3 when the lowest $t$-value is used, even the low $t$-value has a significantly longer execution time, while producing a less accurate estimate, than our simpler algorithm when $M/N$ is very small. Our simpler counting algorithm is asymptotically bounded by $O(\sqrt{N/M})$, the same as the *PEA*-based algorithm, in the number of *controlled-Grover* iterations.

## 2  Quantum circuit for the simpler quantum counting

### 2.1. Quantum circuit for the $k^{th}$ measurement step

The basic quantum circuit consisting of three registers is shown in Fig.1: The first register is the single-qubit *measurement register*. The second is an $n$-qubit *computation register* where the uniform superposition state, $|\psi\rangle$, is the input state to the $2^k$ iterations of the *controlled-Grover* gate. The third register is the workspace for an oracle circuit that is specific to the problem at hand. Note that the Grover operator $G$ is given by a product of the oracle operator $O$ and the *diffusion* operator, $2|\psi\rangle\langle\psi| - I$, as $G = (2|\psi\rangle\langle\psi| - I)O$ [3]. The iterative phase estimation algorithm by Kitaev [4] uses the same circuit as Fig.1 to determine the $k^{th}$ phase bit, with subsequent iterations determining more phase bits until a sufficient resolution for the phase, or an acceptably accurate estimate for $M$, is achieved. In our algorithm, the iteration is used to increase the probability $p_1$ of measuring the state $|1\rangle$ on the measurement register to at least *0.5*. Beyond this point, the probability $p_1$ fluctuates with further increasing the iteration. Our algorithm is in the realm of *amplitude amplification*, not *phase estimation*.

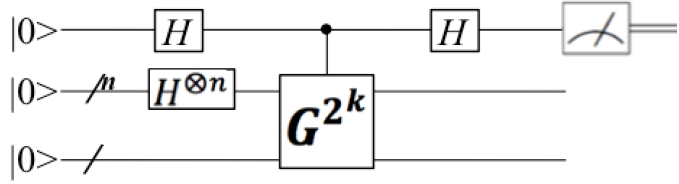

Figure 1. The circuit diagram for our simple quantum counting algorithm at the $k^{th}$ measurement step of the consecutive measurements. As *k* increases starting from zero, the probability of at least *0.5* of measuring */1>* is reached by the final measurement step of $k \leq \lceil log_2 \sqrt{(N/M)} \rceil$.

In order to calculate the measurement probability, we will briefly review the Grover operator and its eigenstates according to ref. [3].

The input to the *Grover* operator *G* is the uniform superposition state *|ψ>* which can be written either in terms of the two eigenstates $|\phi_\pm\rangle$ of the operator *G* or in terms of */χ>*, the normalized state vector representing the uniform superposition of all *marked* states, and */ξ>* representing all *unmarked* states, defined as follows in terms of the $N=2^n$ state vectors */x>* of the search space [3]:

$$|\psi\rangle = \frac{e^{i\theta/2}|\phi_+\rangle + e^{-i\theta/2}|\phi_-\rangle}{\sqrt{2}} = \cos\frac{\theta}{2}|\xi\rangle + \sin\frac{\theta}{2}|\chi\rangle \quad (1)$$

where

$$|\chi\rangle \equiv \frac{1}{\sqrt{M}} \sum_{x,f(x)=1} |x\rangle; \quad |\xi\rangle \equiv \frac{1}{\sqrt{N-M}} \sum_{x,f(x)=0} |x\rangle$$

$$|\phi_\pm\rangle \equiv \frac{|\xi\rangle \mp i|\chi\rangle}{\sqrt{2}}; \quad G|\phi_\pm\rangle = e^{\pm i\theta}|\phi_\pm\rangle; \quad \sin\frac{\theta}{2} \equiv \sqrt{\frac{M}{N}} \quad (2)$$

Here, for each state vector */x>* of the search space, the oracle function returns *f(x)=1* if */x>* is *marked* and *f(x)=0* if */x>* is *unmarked*.

Using the above definitions and relations, we can readily obtain the measurement probability of the first register in Fig.1. After the *controlled-$G^K$* gate where $K=2^k$ and the *Hadamard* gate, the state just before the measurement is (ignoring the state vector of the oracle workspace, the third register, which is */0>*):

$$\cos\frac{K\theta}{2}|0\rangle\left(\cos\frac{(K+1)\theta}{2}|\xi\rangle + \sin\frac{(K+1)\theta}{2}|\chi\rangle\right)$$
$$+ \sin\frac{K\theta}{2}|1\rangle\left(\sin\frac{(K+1)\theta}{2}|\xi\rangle - \cos\frac{(K+1)\theta}{2}|\chi\rangle\right) \quad (3)$$

The amplitude of the $|1\rangle$ state of the measurement register can be increased by increasing the iteration, $K$, hence the *amplitude amplification*. Upon measuring the first register, the probabilities of measuring $0$ and $1$ are, respectively,

$$p_0 = \cos^2 \frac{K\theta}{2}, \quad p_1 = \sin^2 \frac{K\theta}{2} \qquad (4)$$

This may be written, using $K=2^k$ and $\theta=2\pi\phi$, as

$$p_0(k) = \cos^2(\pi 2^k \phi), \quad p_1(k) = \sin^2(\pi 2^k \phi), \quad 0 \leq \phi < 1 \qquad (5)$$

Here, the probabilities are written as a function of the *Grover* iteration parameter $k$. We may also regard $k$ as a parameter representing the $k^{th}$ measurement step. For $k=0$, we have

$$p_0(0) = \cos^2(\pi\phi) = \frac{N-M}{N}, \quad p_1(0) = \sin^2(\pi\phi) = \sin^2 \frac{\theta}{2} = \frac{M}{N} \qquad (6)$$

Although we benefit the most from this simpler algorithm for problems with a very small $M/N$ value, the algorithm works for any $M/N$ value as long as $M/N<½$. If $M/N≥½$, then we simply double the size of the search space to $2N$ states by adding one more qubit to the computation register (the second register in Fig.1), making it a register with $n+1$ qubits. So, we will discuss the case with $M/N<½$ only and we focus mostly on a very small $M/N << 1$.

*2.2. Discussion of the final measurement step*

The probability $p_1$ depends on $M/N$ via $p_1(0) = \sin^2(\theta/2) = M/N$. For a very small $M/N$, where the measurement probability is very small, we employ the *controlled-Grover* iteration to boost the probability $p_1$. This tactic is similar to the Grover search algorithm where the Grover iteration is employed to increase the amplitudes of the *marked* states [1].

For $\theta = 2\pi\phi$ and $0\leq\phi<1$, let us express $\phi$ as a *binary fraction*:

$$\phi = 0.\phi_1\phi_2\phi_3... \qquad \text{(binary fraction)} \qquad (7)$$

Since $\sin^2(\pi\phi) = M/N < ½$, it follows that $\phi < 2^{-2}$, and hence $\phi_1=0$ and $\phi_2=0$. Therefore, $\phi = 0.00\phi_3\phi_4\phi_5...$ for $M < ½N$.

**Proposition**: For a very small $M/N$ value, the probability of measuring $1$ on the measurement register in Fig.1 is $p_1(k)$ where $p_1(k)=\sin^2(\pi 2^k\phi)$ after iterating the *controlled-Grover* gate for $2^k$ times. The consecutive measurement *starts* at $k=0$ and *stops* at the first measurement in which $p_1(k)$ reaches or exceeds $0.5$. The condition $0.5\leq p_1(k)\leq 1$ is satisfied when $k=\lceil log_2 \sqrt{(N/M)} \rceil$ or when $k=\lceil log_2 \sqrt{(N/M)} \rceil-1$ depending on the $M/N$-value. Hence, the maximum possible number of measurement steps required is $\lceil log_2 \sqrt{(N/M)} \rceil$.

**Proof**: For a small $M/N$ value (*i.e.*, $M/N << 1$), we approximate $\theta/2 \approx \sqrt{(M/N)}$ and $\phi \approx (1/\pi)\sqrt{(M/N)}$. After the *controlled-Grover* operation is iterated for $K=2^k$ times, the probability $p_1(k)$ is $\sin^2(\pi 2^k\phi)$ according to eq.(5). The maximum or optimal number $2^k$ of the *controlled-Grover* iteration for estimating $M/N$ is chosen such that

$$2^{-2} \leq 2^k \phi \leq 2^{-1} \rightarrow 0.5 \leq p_1(k) \leq 1 \qquad (8)$$

The $k$-value satisfying this condition is the first $k$ value that makes $p_1(k)$ greater than or equal to $0.5$ when $k$ is raised consecutively starting from zero. Each increment in $k$ amounts to a left-shift by 1 bit of every $\phi_j$ for $j \geq 1$. The $k$-value satisfying the above condition eq.(8) is

$$k = \lceil (n-log_2 M)/2 \rceil \qquad (9)$$

or one less, as can be shown as follows: Let $k$ be the integer given by eq.(9). Let us define $r$, a nonnegative value (*double*) by:

$$r \equiv \left\lceil \frac{n-log_2 M}{2} \right\rceil - \frac{n-log_2(M)}{2} = k - log_2 \sqrt{\frac{N}{M}}, \; 0 \leq r < 1 \qquad (10)$$

Let us also define a constant $a$ by $1/\pi \equiv 2^{-2+a}$ where $a \approx 0.3485$. Then, for $M/N << 1$,

$$2^k \phi \approx \frac{2^k}{\pi} \sqrt{\frac{M}{N}} = \frac{1}{\pi} 2^{k-(k-r)} = 2^{-2+a+r} \qquad (11)$$

Hence, if $a+r \leq 1$, then the condition eq.(8) is satisfied with the $k$-value given by eq.(9). If the $M/N$-value is such that $a+r > 1$, then the $k$-value of eq.(9) minus $1$ is automatically chosen because the measurement stops at the first $k$-value such that $p_1(k)$ is raised to or past $0.5$. Therefore, the condition in eq.(8) is satisfied in at most $\lceil log_2 \sqrt{(N/M)} \rceil$ measurement steps. □

In practice, one could require a value less than $0.5$ for $p_1(k)$ to stop the iteration, such as $p_1(k) \geq p_{min}=0.15$, then the algorithm may halt at a lower $k$-value and still produce a good estimate for $M$. A classical post-processing is used to calculate $M$ from the probabilities, $p_0(k)$ and $p_1(k)$, found in the *final*, $k^{th}$ iteration step. One could also determine the phase angle $\theta$ of the Grover eigenvalue for use in obtaining the optimal number of Grover iterations $R$ in the Grover search algorithm as shown below.

From the measured probabilities $p_0(k)$ and $p_1(k)$ of $0$ and $1$, respectively, we define:

$$p(k) \equiv p_0(k) - p_1(k) = cos^2(\pi 2^k \phi) - sin^2(\pi 2^k \phi) = cos(2^k \theta) \qquad (12)$$

where $\theta = 2\pi\phi$ and $k$ is the $k$-value of the *final* iteration step of the algorithm. The probabilities, $p_0(k)$ and $p_1(k)$, used in the estimation of $M$ come *only* from the *final* iteration step. The classical post-processing can use one of the following two formulas:

Classical post-processing formulae-1:

$$\theta = 2^{-k} cos^{-1} p(k); \; M = N sin^2 \frac{\theta}{2} \qquad (13)$$

Classical post-processing formulae-2:

$$p(k-1) = \sqrt{\frac{1+p(k)}{2}} \rightarrow \text{iterate } k \text{ times to obtain } p(0)=cos\theta. \qquad (14)$$

$$M = N\frac{1-p(0)}{2}, \quad \theta = \cos^{-1}p(0) \tag{15}$$

From the measured Grover angle $\theta$, the optimal number $R$ of the Grover iterations in the quantum search algorithm [1, 3] is

$$R = \left\lceil \frac{\frac{\pi}{\theta}-1}{2} \right\rceil \tag{16}$$

**Algorithm**: *Simpler Quantum Counting*
**Inputs**: (1) A single-qubit register for measurement, initialized to $|0\rangle$; An $n$-qubit register for computation in a uniform superposition state; and a multi-qubit register for oracle workspace, initialized to $|0\rangle$. (2) The function oracle for the problem.
**Outputs**: Approximate value of $M$
**Runtime**: $O(\sqrt{N/M})$ calls to *controlled-Grover* gate; $\lceil log_2 \sqrt{N/M} \rceil$ measurement steps. Succeeds with probability $O(1)$.
**Procedure**:
  1. **Initialize:** $p_1=0$, $k=-1$
  2. **while** ($p_1<0.5$) #loop stops when $p_1 \geq 0.5$
  3.     $k=k+1$
         #*perform the circuit in Fig.1 as the $k^{th}$ measurement step*
  4.     Reset states to $|0\rangle$; prepare as in the circuit in Fig.1 (*i.e.*, apply *Hadamard* gates)
  5.     Perform *controlled-$G^{2^k}$* in the circuit in Fig.1
  6.     Apply *Hadamard* to the *first* register; measure $p_1=p_1(k)=Pr(|1\rangle)$ and $p_0(k)=Pr(|0\rangle)$
  7. **Classical post-processing**:  #*only the final step above, with $p_1(k) \geq 0.5$, is used below.*
  8.     $p(k)=p_0(k)-p_1(k)$ →
  9.     $p(k-1) = \sqrt{\frac{1+p(k)}{2}}$ → iterate $k$ times to obtain $p(0)=\cos\theta$.
  10. **Return** an approximate value of $M \approx N(1-p(0))/2$

## 3 Comparison of the two algorithms by simulation

Here, we compare our algorithm against the *QFT*-based phase estimation for the quantum counting [4]. Our algorithm executes the measurements in sequence in at most $\lceil log_2 \sqrt{N/M} \rceil$ measurement-steps when $p_1(k) \geq 0.5$ is achieved. This means that the $(k+2)^{th}$ bit is the most significant *1-bit* of $\phi$. We call this algorithm *simple* in the sense that the measurement register consists of a single qubit where the measurement probability is split between only two states, $|0\rangle$ and $|1\rangle$, and there exists a clear criterion for a successful measurement (*i.e.*, $p_1(k) \geq 0.5$). The total number of *controlled-Grover* iterations needed for a successful measurement can be substantially smaller, and the quantum circuit significantly narrower than the *PEA*-based counting algorithm, especially when the *M/N*-value is small.

In our simpler counting algorithm, the total number of *controlled-Grover* iterations in all measurement steps up to the $k^{th}$ step (starting at $k=0$) is $2^0 + 2^1 + ... + 2^k = 2^{k+1}-1$. The *PEA*-based algorithm with *t*-qubits in the measurement register has a total of $2^t-1$ controlled-Grover iterations. Thus, all iteration steps up to the $k^{th}$ step of our simpler algorithm and the *PEA*-based algorithm with $k+1$ qubits in the measurement register would have the same number of *controlled-Grover* iterations. The $k^{th}$ measurement step alone of the simpler algorithm and the *PEA*-based algorithm with $k$-qubits in the measurement register would have nearly the same circuit depth. Also, our simpler algorithm repeats the state initialization and preparation for each step, while the *PEA*-based algorithm performs an inverse Fourier transform on the measurement register before measurement.

Before comparing the two counting algorithms by simulation, we first summarize below the *PEA*-based algorithm according to ref.[3]. The *PEA*-based quantum counting circuit is shown in Fig.2.

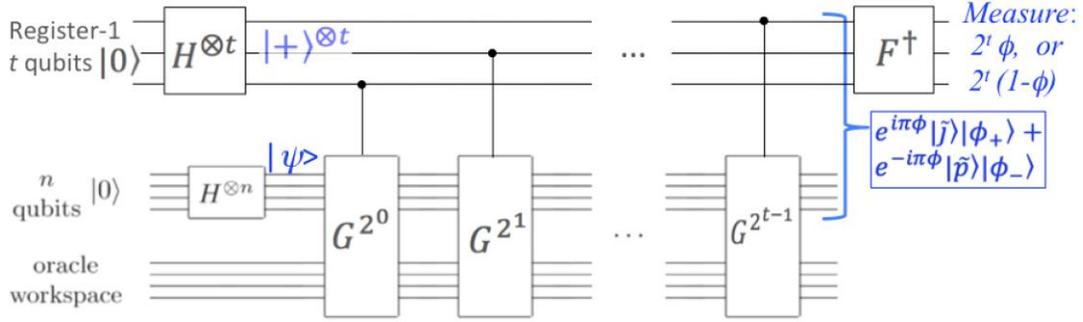

Figure 2. Quantum circuit for the counting algorithm based on the phase estimation algorithm [3].

The number of qubits $t$ of the first register is determined from the desired measurement accuracy $m$ and the probability $1-\varepsilon$ of a successful measurement [3].

$$t = m + log_2(2+1/2\varepsilon) \qquad (17)$$

This number $t$ should be large enough for the estimated-$M$ to be sufficiently close to the actual value with a high measurement probability. For $M<\frac{1}{2}N$ where $\phi_1=0=\phi_2$, we get from the measurement results of $\phi$:

$$M = N\ sin^2(\pi(\phi_3 2^{-3} + \phi_4 2^{-4} +..+\phi_t 2^{-t})) \qquad (18)$$

Here, $\phi_j$ is the measured $j^{th}$ bit of the *register-1* in Fig.2. (*Note*: if $M \geq \frac{1}{2}N$, then we simply double the search space by adding another qubit to the $2^{nd}$ register and make the search space $N=2^n \rightarrow 2N = 2^{n+1}$ by replacing $n$ by $n+1$ in all equations below.)

The quantum state just before the inverse Fourier transform $F^\dagger$ is a superposition of the two Grover eigenstates $|\phi_+\rangle$ and $|\phi_-\rangle$ entangled with the states $|\tilde{j}\rangle$ and $|\tilde{p}\rangle$ which are the Fourier transformed states of $|j\rangle=|j_1 j_2..j_t\rangle$ and $|p\rangle=|p_1 p_2..p_t\rangle$, respectively, where $j$ and $p$ are defined by the following relations:

$$\phi \equiv \frac{j}{2^t},\ 1-\phi \equiv \frac{p}{2^t},\ 0 \leq j,p < 2^t \qquad (19)$$

The state just before $F^\dagger$ is

$$e^{i\theta/2}|\tilde{j}\rangle|\phi_+\rangle + e^{-i\theta/2}|\tilde{p}\rangle|\phi_-\rangle = e^{i\theta/2}|\widetilde{2^t\phi}\rangle|\phi_+\rangle + e^{-i\theta/2}|\widetilde{2^t(1-\phi)}\rangle|\phi_-\rangle \qquad (20)$$

After applying the inverse *QFT*, $F^\dagger$, we measure all *t* qubits of *register-1* and obtain either *j* or *p* to a *t*-bit precision, thereby giving an approximate value of $\phi$ or *1-$\phi$*. They both yield the same *M* value:

$$M = 2^n \sin^2(\pi(1-\phi)) = 2^n \sin^2(\pi\phi) \tag{21}$$

Implementation of various quantum algorithms for execution on a 5-qubit IBM quantum computer can be found in ref. [6]. The same implementation approach can be used for simulating a larger quantum circuit.

We applied the two counting algorithms to estimate the number of maximal cliques in a graph where the uniform superposition state $|\psi\rangle$ includes *M* maximal clique states [7]. These maximal clique states are the *marked* items in the search space of $2^n$ states. An oracle circuit for finding the maximal cliques from an adjacency matrix data for a graph of an arbitrary size was given in ref. [7]. A graph with three nodes (*n=3*) was considered here because the oracle requires additional $2n^2$ qubits for data and ancilla [7]. Hence, the largest graph that we could simulate using an available simulator (the 32-qubit IBM simulator [8]) was a graph with only three nodes. This example provides a comparison between the two counting algorithms with a not so small *M/N* value of 0.125, 0.25 or 0.375.

Figures 3 and 4 show the simulation results for the maximal cliques where all nodes are connected to each other, hence only one maximal clique in the graph, *M=1*.

Figure 4 shows the simulation results from the *PEA*-based algorithm with three control bits along with a plot of the measured *M* and their respective probabilities as a function of *t*. According to eq.(21), the two states whose binary fractions ($\phi$) add to unity represent the same *M*-value. For example, the states 001 and 111 give the same $\sin^2(\pi\phi)=\sin^2(\pi(1-\phi))=M/N$ and hence their probabilities are added for the given measured *M* in Fig.4 (right). In this case, the sum of probabilities for the states 001 and 111 is 98%. From this result, the measured *M*-value is $2^3 \times 0.1463=1.17$, which is plotted in the second graph in Fig.4. In the case of *t=4*, for each of the 1024 measurements performed, the *controlled-Grover* operator would iterate $2^t-1=15$ times.

The example in Fig.3 illustrates the usefulness of our simple quantum counting algorithm. If the probability $p_1$ is too small due to a small *M/N* value, we simply iterate the *controlled-Grover* operator to boost the probability $p_1$, in a manner similar to the quantum search algorithm which uses the *Grover* iteration to enhance the search probability [1]. In the quantum search algorithm, each time the *Grover* operator is iterated the measurement probabilities are changed uniformly across all *marked* states by a factor, and across all *unmarked* states by another factor. In the quantum search, the amplitude of marked states increases monotonically up to a certain iteration, and then decreases upon further increasing the iteration. In our simpler counting, the probability $p_1(k)$ increases monotonically with increasing *k* until $p_1(k) \geq 0.5$, and beyond this point, $p_1(k)$ fluctuates (depending on the lower significant bit values of $\phi$) with further increasing *k*. In the simple counting algorithm, only two states split the probability, making it simpler and more reliable.

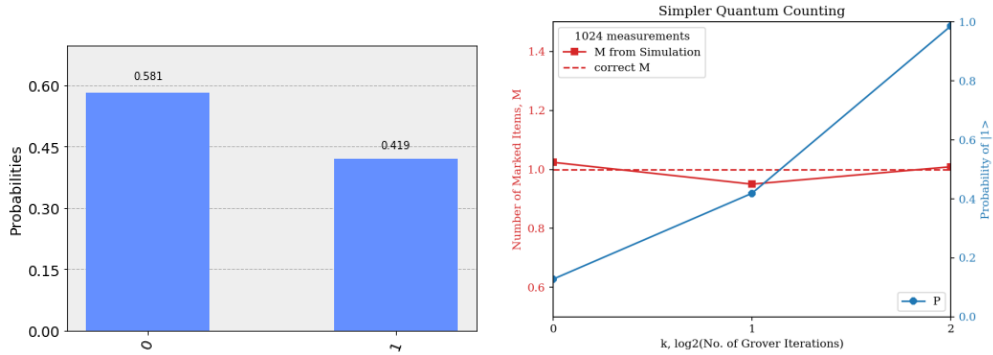

Figure 3. Simple Quantum Counting: The simulation result for $M/N = 1/2^3 = 0.125$, with $K=2^k$ *controlled-Grover* iterations with $k=1$ (*left*), and the measured $M$ vs. $k$ with $p_1(k)$ (*right*). Each step had 1,024 measurements. Dashed horizontal line indicates the correct number ($M=1$).

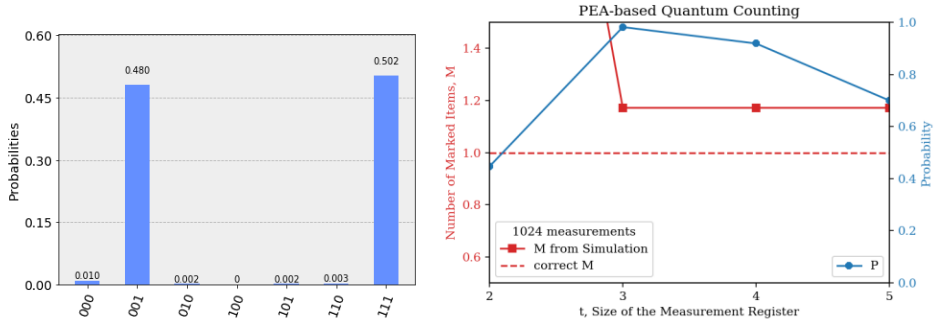

Figure 4. *PEA*-based counting of maximal cliques for $M/N=1/2^3$: The simulation results with $t=3$ qubits in the measurement register (*left*), and the measured number ($M$) of maximal cliques and the highest probability sum of the two phase bits that add to $2^t$ (*right*).

Figures 5 and 6 show the simulation results for the number of maximal cliques with $M=2$ and $M=3$, respectively, in the graph of three nodes. Even for these cases with a large $M/N$ value (0.25 and 0.375, respectively), the estimated $M$ value is closer to the correct value for the simpler algorithm than for the *PEA*-based one.

In the *PEA*-based counting, the measurement probability is shared between the two equivalent phases, $\phi$ and $1-\phi$, making each probability only about one half of the probability if only one phase angle represented the given $M$ value.

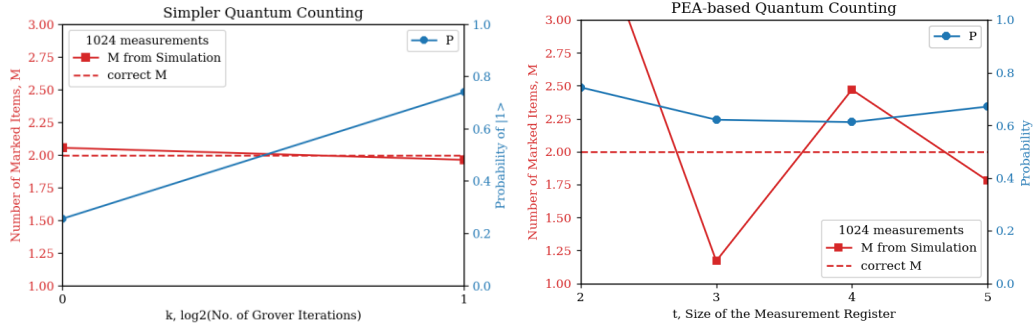

Figure 5. For $M=2$ and $N=2^3$, the estimated number $M$ of maximal cliques by the simpler counting algorithm ($1^{st}$ graph) and by the *PEA*-based counting algorithm ($2^{nd}$ graph) along with their respective measurement probabilities. For the *PEA*-based counting algorithm, the measurement probability is the sum of the highest probabilities of the two bit-trains that add to $2^t$.

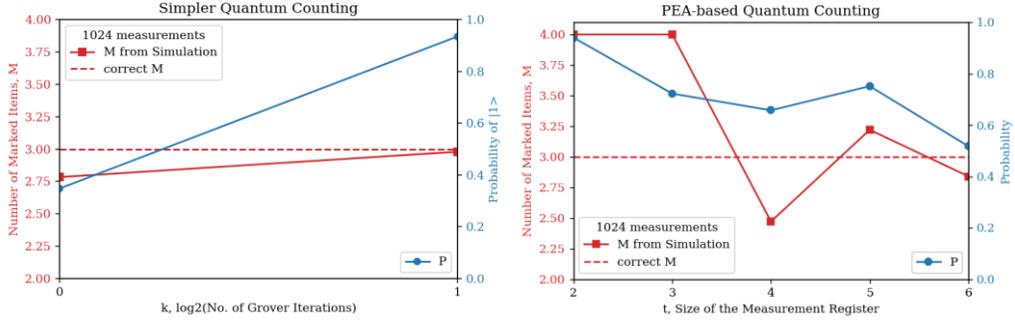

Figure 6. For $M=3$ and $N=2^3$, the measured $M$ values from the simpler counting algorithm ($1^{st}$ graph) and the *PEA*-based counting algorithm ($2^{nd}$ graph) along with their respective measurement probabilities.

Our simpler counting algorithm is required to terminate at the first measurement step where $p_1(k) \geq 0.5$, which is satisfied at $k = \lceil log \sqrt{(N/M)} \rceil$ according to

$$2^k \phi = 2^{k-log\sqrt{(N/M)}}/\pi = 1.01\ldots_2 2^{k-2-log\sqrt{(N/M)}} \geq 2^{-2} \qquad (22)$$

On the other hand, in the *PEA*-based algorithm, the relation, $\phi = 1.01\ldots_2 2^{-2-log\sqrt{(N/M)}}$, suggests that the least significant bit (*LSB*) will be the only nonzero bit if the measurement (phase) register had $2+\lceil log \sqrt{(N/M)} \rceil$ qubits. This means that the *PEA*-based algorithm requires $t \geq 2+\lceil log \sqrt{(N/M)} \rceil$. This made some simulations already impractical if $M/N < 0.1\%$ (that is, $t \geq 7$) using the 32-qubit IBM simulator [8]. The smallest nonzero $M/N$-value that can be estimated by the *PEA*-based algorithm is $sin^2(\pi/2^t)$. Hence, for a very small $M/N$ value, the number of qubits required for the measurement register is necessarily large.

In order to compare the two algorithms for problems with a very small $M/N$ value, and still be able to perform the simulation on a currently available simulator, we formed a very simple oracle which searches for the integer numbers whose binary bits are *1* at certain specific bit positions in an *n*-bit binary representation of integers between *0* and $2^n$-*1*. This oracle circuit is formed by a *controlled-NOT* gate, $C^j(X)$, where the *j* control bits are the bits that must be *1* and its output controls a *phase-flip gate (Z)* that acts on an ancilla qubit in the state *|1>*, to complete the oracle function. Using the *32-qubit simulator*, we tested up to *n=12* and our simple algorithm had a reasonably short execution time. We also performed the same simulation using the *PEA*-based algorithm as long as the program could be run within a reasonable time (a few hours). The simulation results are presented below for the varying values of *M/N*.

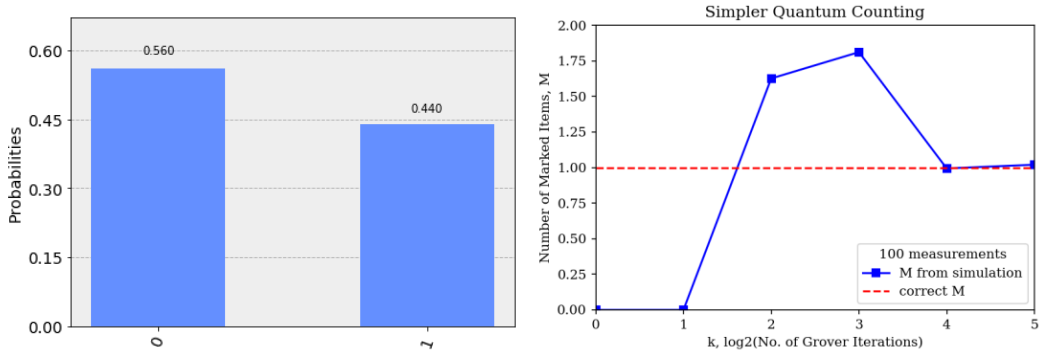

Figure 7. $N=2^9=512$, *M=1*. Simpler counting algorithm: The simulation result for *k=4* (*1st graph*) and the measured *M vs. k* for 100 measurements (*2nd graph*) where the red horizontal dashed line indicates the correct *M* value.

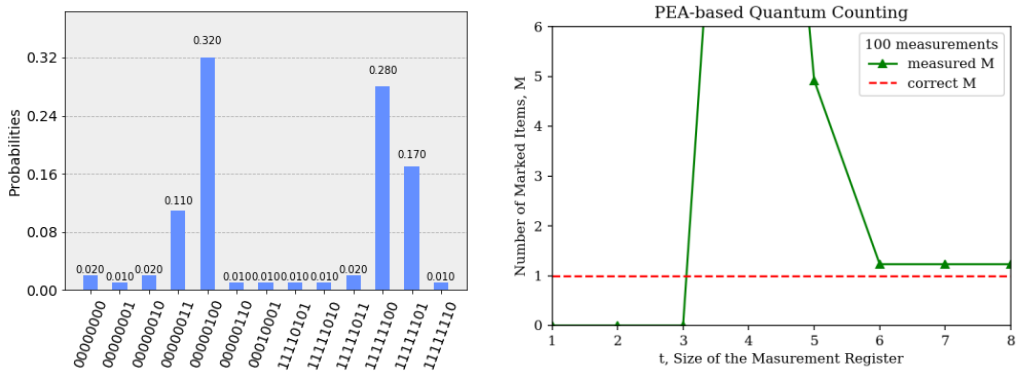

Figure 8. $N=2^9$, *M=1*. *PEA*-based algorithm: The first graph shows simulation result for *t=8*. The highest probability *M*-value is shown *vs. t* in the second graph. The correct *M*-value is indicated by the red horizontal dashed line.

For example, consider integers between *0* and $2^9$-*1*, and assume that the oracle searches for the *9*-bit integers where every bit is *1*. This is the case where $N=2^9$ and *M=1*. The simulation results from the two algorithms are shown in Figures 7 and 8.

In order to test an even smaller *M/N* value, we used $N=2^{12}=4,096$ and a varying *M*. Figures 9-16 show the simulation results from the simpler counting algorithm as a function of the *controlled-Grover* iteration parameter *k* at constant *100* measurements (*1st* graphs), and as a function of the number of measurements (*2nd* graphs) at the maximum *k* value from the first graph (*i.e.*, the first *k*-value at which $p_1(k)$ is at least *0.5*). For *n=12* and for some smaller *M* values, the *PEA*-based algorithm did not yield any result within a reasonable run time (a few hours) when *t* was greater than *6* (*t>6*) (Figures 9-11) and thus *t* was limited to a maximum *6* for the $N=2^{12}$ case, and the returned results are shown in the last graphs in Figures 12-16. The simulation results with the simpler algorithm show for all cases that the measured *M*-value is a good estimate of the correct *M* value for the last two *k*-values in the plot (corresponding to *$p_1(k)$ just below and just above 0.5*). This can be seen in the first graphs in the Figures 9 through 16.

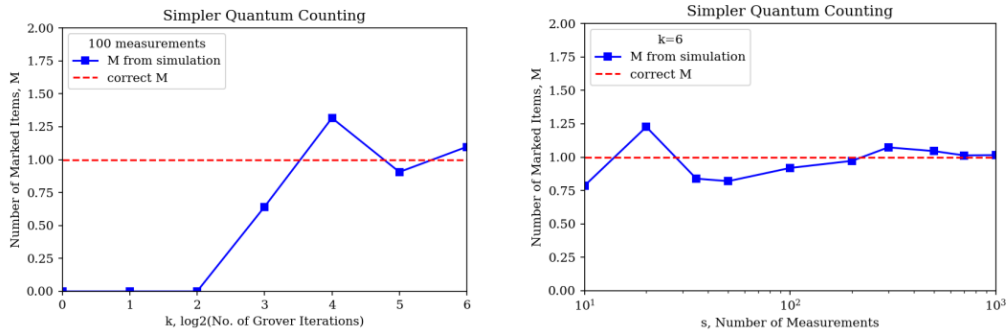

Figure 9. *$N=2^{12}=4096$, M=1, M/N=0.024%*. Measured *M* from the simple counting algorithm as a function of the *controlled-Grover* iterations (*1st graph*) and as a function of the number of measurements *s* at *k=6* (*2nd graph*). The simulation yields a reliable result for *k=5 ($p_1(k) < 0.5$)* and *k=6 ($p_1(k) > 0.5$)* with *100* measurements.

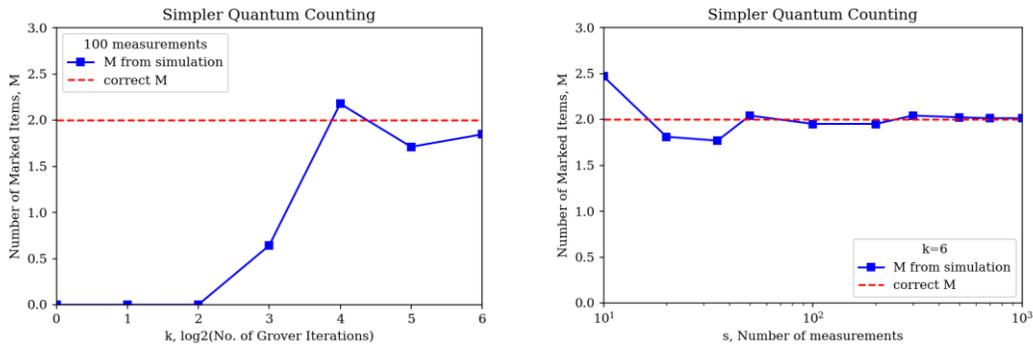

Figure 10. *$N=2^{12}$, M=2, M/N=0.049%*. Simpler counting: measured *M* vs. *k* (*1st graph*); measured *M* at *k=6* vs. the number of measurements (*2nd graph*).

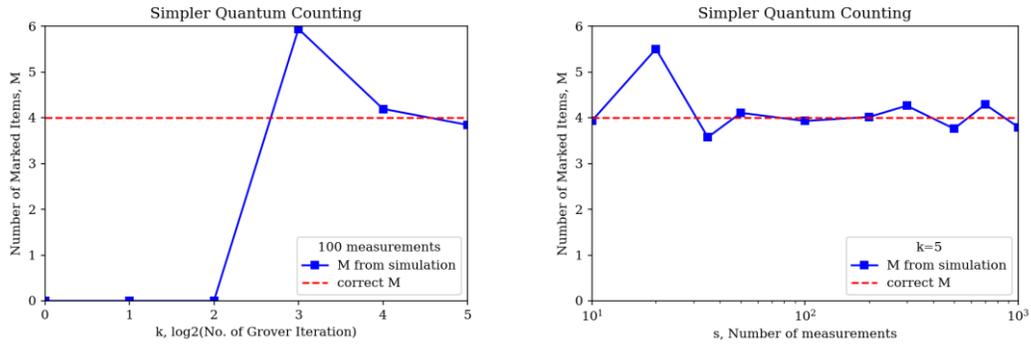

Figure 11. $N=2^{12}$, $M=4$, $M/N=0.098\%$. Simpler counting: measured $M$ vs. $k$ ($1^{st}$ graph); measured $M$ at $k=5$ vs. the number of measurements ($2^{nd}$ graph).

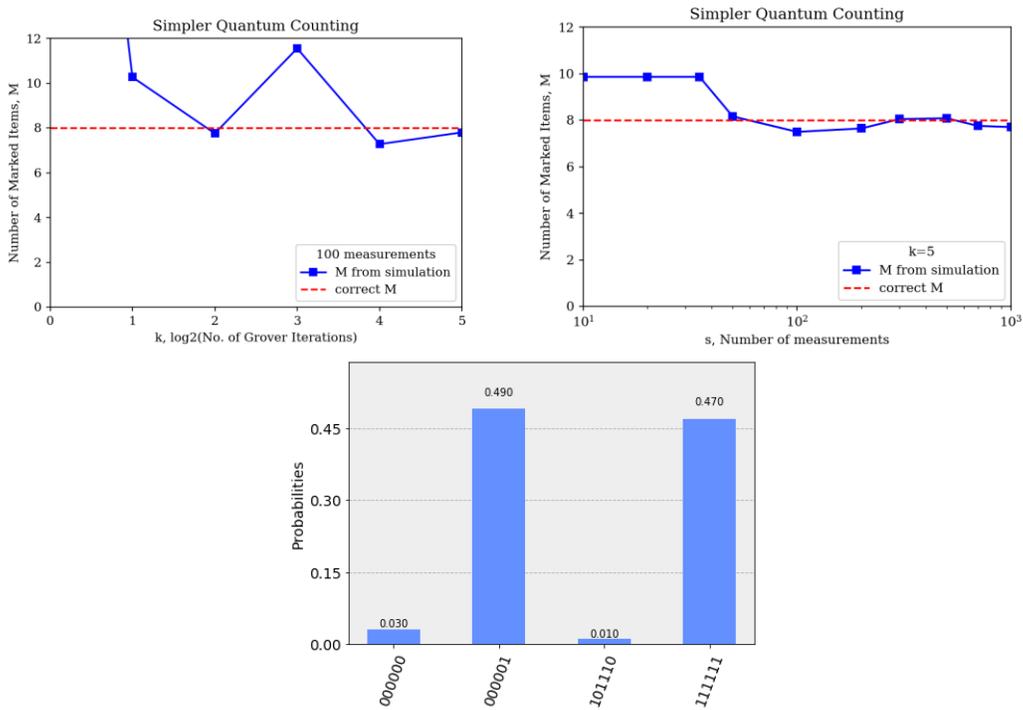

Figure 12. $N=2^{12}$, $M=8$, $M/N=0.195\%$. The simulation results by the simpler counting algorithm (first two graphs) and by the *PEA*-based algorithm with $t=6$ measurement qubits, yielding $M=9.86$ (last graph).

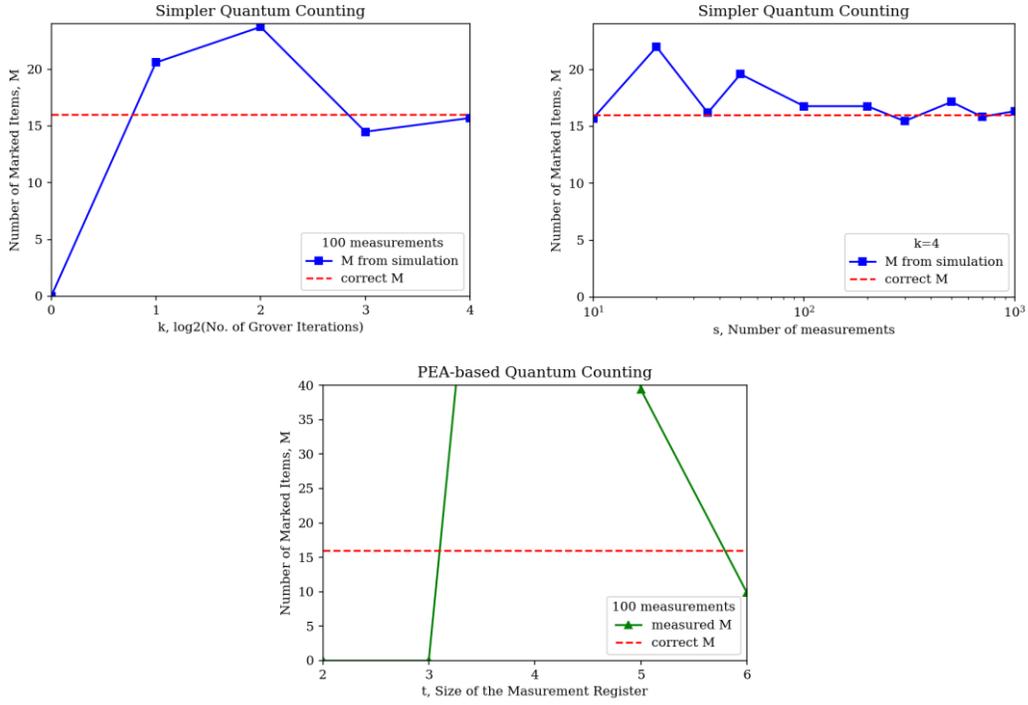

Figure 13. $N=2^{12}$, $M=16$, $M/N=0.391\%$. Simpler counting algorithm: measured *M* vs. *k* and measured *M* at *k=4 vs.* number of measurements (first two graphs); and simulation result by the *PEA*-based algorithm (*3$^{rd}$ graph*).

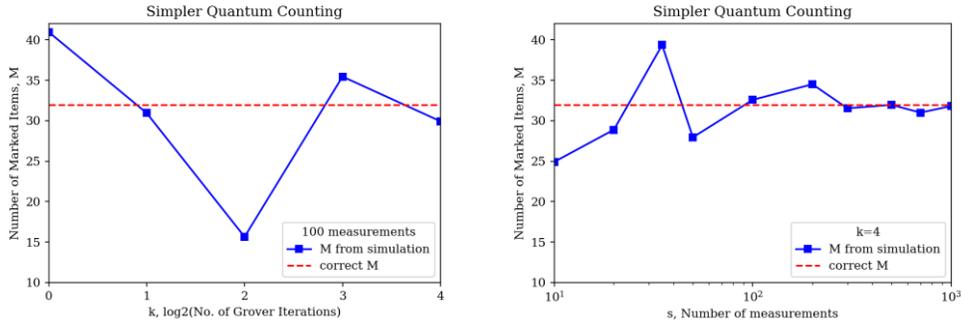

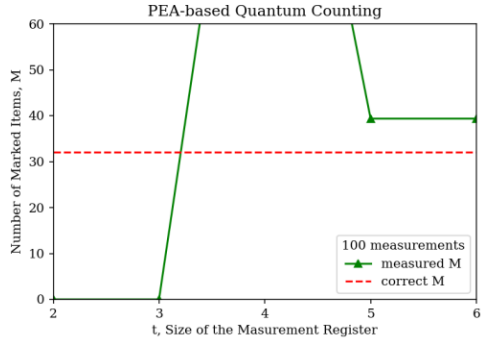

Figure 14. $N=2^{12}$, $M=32$, $M/N=0.781\%$. Simpler counting algorithm: measured *M* vs. *k* and measured *M* at *k=4 vs.* number of measurements (first two graphs); and the result by the *PEA*-based algorithm (*3$^{rd}$ graph*).

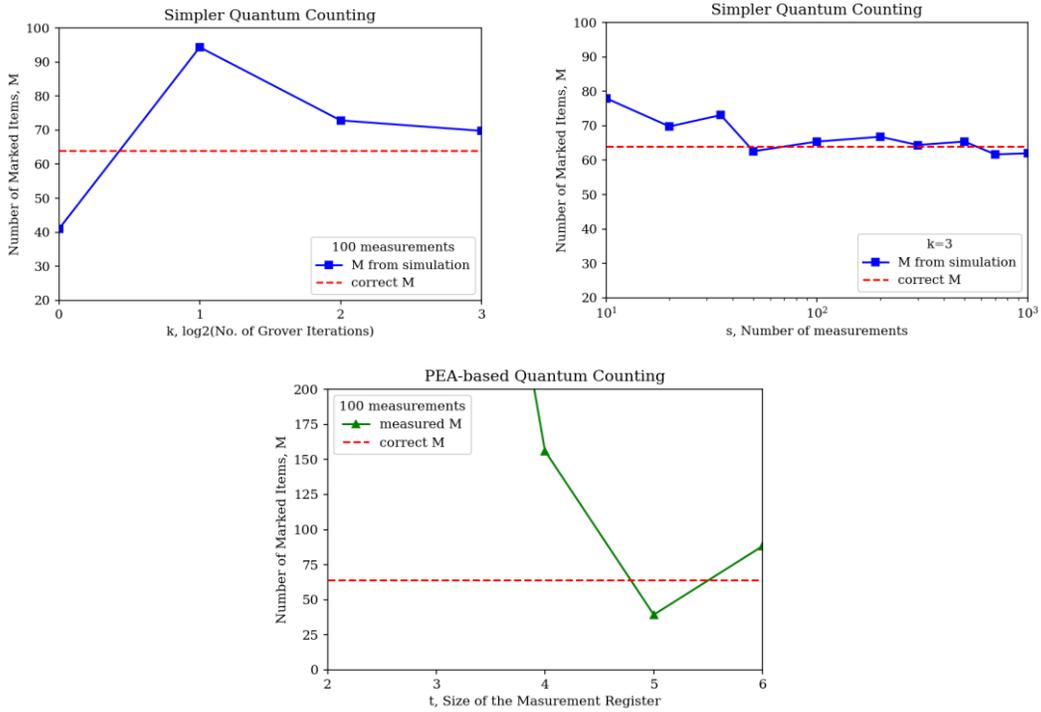

Figure 15. $N=2^{12}$, $M=64$, $M/N=1.563\%$. Simpler algorithm: measured *M* vs. *k* and measured *M* at *k=3 vs.* the number of measurements (first two graphs); and the simulation result by the *PEA*-based algorithm (*3$^{rd}$ graph*).

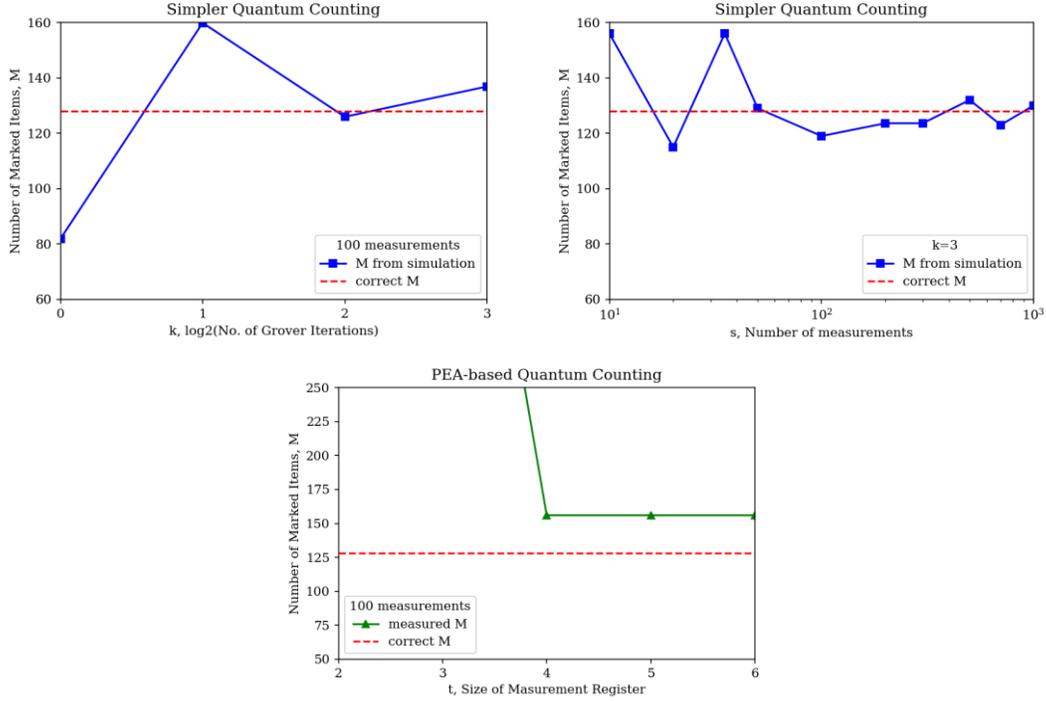

Figure 16. $N=2^{12}$, $M=128$, $M/N=3.125\%$. Simpler algorithm: measured $M$ vs. $k$ and measured $M$ at $k=3$ vs. number of measurements (first two graphs); and the simulation result by the *PEA*-based algorithm ($3^{rd}$ graph).

## 4  Summary

In summary, we presented a simpler quantum counting algorithm based on amplitude amplification. This algorithm is more reliable and practical, than the *PEA*-based algorithm, especially for problems with a small *M/N*. The measured *M*-value by this simpler counting algorithm gave a good estimate for the last measurement step where $p_1(k)$ is *0.5* or greater. For the last ($k^{th}$) step where $p_1(k) \geq 0.5$, the simpler algorithm produced a better estimate than the best-case result by the *PEA*-based algorithm even with the maximum circuit width allowed and the maximum run time practicable in currently available quantum simulators. The result from the final iteration step with $p_1 \geq 0.5$ is used to estimate *M*.

**Note added at the time of proof**: Shortly after this article was submitted, Aaronson and Rall reported a *QFT-free* quantum approximate counting algorithm with rigorous analyses and with an excellent introduction on the state of quantum approximate counting algorithms, pointing out respective features [9]. I gratefully acknowledge many helpful discussions with Scott Aaronson and Patrick Rall.